# Perfectly-reflecting guided-mode-resonant photonic lattices possessing Mie modal memory

Yeong Hwan Ko[1], Nasrin Razmjooei[1], Hafez Hemmati and Robert Magnusson*

Department of Electrical Engineering, University of Texas at Arlington, Arlington, Texas 76019, USA
*Author to whom correspondence should be addressed: magnusson@uta.edu
[1]Equal first authors

**ABSTRACT**

Resonant periodic nanostructures provide perfect reflection across small or large spectral bandwidths depending on the choice of materials and design parameters. This effect has been known for decades, observed theoretically and experimentally via one-dimensional and two-dimensional structures commonly known as resonant gratings, metamaterials, and metasurfaces. The physical cause of this extraordinary phenomenon is guided-mode resonance mediated by lateral Bloch modes excited by evanescent diffraction orders in the subwavelength regime. In recent years, hundreds of papers have declared Fabry-Perot or Mie resonance to be basis of the perfect reflection possessed by periodic metasurfaces. Treating a simple one-dimensional cylindrical-rod lattice, here we show clearly and unambiguously that Mie resonance does not cause perfect reflection. In fact, the spectral placement of the Bloch-mode-mediated zero-order reflectance is primarily controlled by the lattice period by way of its direct effect on the homogenized effective-medium refractive index of the lattice. In general, perfect reflection appears away from Mie resonance. However, when the lateral leaky-mode field profiles approach the isolated-particle Mie field profiles, the resonance locus tends towards the Mie resonance wavelength. The fact that the lattice fields "remember" the isolated particle fields is referred here as "Mie modal memory." On erasure of the Mie memory by an index-matched sublayer, we show that perfect reflection survives with the resonance locus approaching the homogenized effective-medium waveguide locus. The results presented here will aid in clarifying the physical basis of general resonant photonic lattices.

**Introduction**

Periodic arrays of dielectric nanostructures support remarkable resonance effects as incident light couples to leaky Bloch-type modes.[1-8] At resonance, there appears resonant reflection where the reflectance approaches 100% across a particular spectral bandwidth for subwavelength periods. Thirty years ago, the term "guided-mode resonance (GMR)" was coined to communicate clearly the fundamental physics governing the effect.[9] In earlier literature, authors sometimes referred to these effects as "anomalous reflection".[1-3] In recent literature, traditional periodic structures including GMR resonance devices are commonly called photonic crystals, metasurfaces, or metamaterials. It is clear that this class of resonance devices can possess one-dimensional (1D) or 2D lateral spatial modulation, or periodicity, as the resonance physics is not dependent on the type of periodicity in any fundamental way. The most expeditious route towards clear understanding is to model the canonical 1D lattice as all of the main properties reside therein. This is our approach here.

Fabry-Perot (FP) resonance occurs via reflections between parallel planes possessing refractive-index discontinuities and is typically associated with thin films. Mie resonance occurs via similar reflections but between nonparallel planes and is generally associated with isolated cylindrical and spherical particles. Taking a glass particle with refractive index n=1.5 as an example, FP and Mie resonances in air would manifest based on ~4% reflection at each interface. Thus, intuitively, one would not expect high reflection off that particle, or even an array of such particles, via the FP or Mie resonance mechanism at any wavelength. In complete contrast, a periodic glass-particle lattice supporting guided-mode or leaky-mode resonance generates ~100% plane-wave reflectance at the resonance wavelengths.[1-10]

The relevance of this discussion in the present context is that, in recent years, hundreds of scientific papers have declared FP or Mie resonance as the basis of the perfect reflection possessed by periodic 1D and 2D metasurfaces. We can cite only a few representative examples here.[11-17] In these publications, the prior works with a plethora of relevant results on perfect reflection are rarely mentioned. Consequently, the true physical mechanism behind perfect reflection that is grounded in lateral leaky Bloch modes and evanescent-wave resonance excitation is not understood or ignored. Objections to the FP resonance picture[18-19] and the Mie resonance explanation[20] have been published previously.

The papers that claim Mie resonance as origin of high reflection generally discuss resonance properties of isolated particles in some detail and then proceed to periodic or quasi-periodic arrays and their reflection properties. Reflectance plots may label spectral locations of the electric and magnetic Mie dipoles implying that these provide the operative mechanisms supporting the spectrum. The connection of the reflectance spectra with the dominant mechanisms provided by the periodic lattice is not explained. This lack of clarity has led to a plethora of works in the literature claiming that the resonances observed in the isolated particles literally cause perfect reflection. Thus, we believe that a clear and unambiguous distinction is needed which is provided herein.

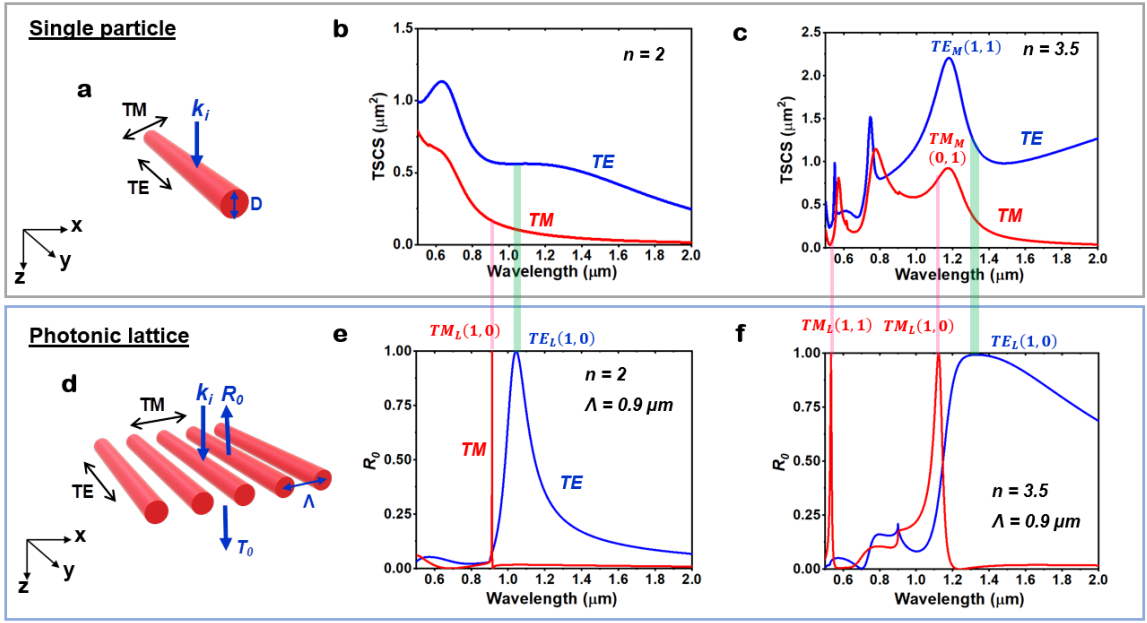

**Fig. 1 | Comparison of single-particle resonance spectra (a-c) and lattice-resonance spectra (d-f). a**, The particle chosen is an infinite circular cylinder with diameter D and refractive index n placed in air. **b**, FDTD-computed TSCS spectra under TE and TM polarized light with n=2. **c**, TSCS spectra for n=3.5 where the Mie resonance peaks are labeled $TE_M(j,l)$ or $TM_M(j,l)$ by the azimuthal mode number $j$ and the radial mode number $l$. **d**, A photonic lattice arrayed by the elemental cylinder in **a**. With a representative period ($\Lambda$), the zeroth-order reflectance ($R_0$) spectra are calculated by RCWA. **e**, $R_0$ spectra under TE and TM polarization for n=2. The resonance peaks are labeled $TE_L(m,v)$ or $TM_L(m,v)$ where $m$ denotes the evanescent diffraction order and $v$ the waveguide mode. **f**, $R_0$ spectra for n=3.5. In this example, there is no correlation between the Mie resonance peaks and the lattice resonance peaks.

The true physics of resonant optical lattices has been presented in many prior works. For example, the coupled-wave equations governing wave propagation in periodic films were shown to convert to the wave equation for a slab waveguide in the limit of small modulation.[4] Subsequent rigorous calculations strongly associated the resonance wavelengths with waveguide modes in the corresponding waveguide; hence, the descriptive terminology "guided-mode resonance." Rosenblatt et al. presented analytic and numerical models detailing the occurrence of lateral leaky modes and establishing explicit phase relations supporting resonance energy transport in reflection.[6] Niraula et al. addressed the mode-coupling mechanisms involved in realizing resonant bandpass filters.[21] There, the roles of the lateral evanescent diffraction orders in sculpting the observed spectral characteristics are clearly presented. In particular, a mode excited by a second evanescent diffraction order can attain a dominant strength and generate efficient transmittance. As an aside, we note that this version of the guided-mode resonance effect is sometimes, somewhat inaccurately, called "electromagnetically induced transparency" on account of the full transmission of a lossless lattice simultaneously providing low transmittance sidebands via its lateral modal content. One of the major shortcomings of the local FP or Mie resonance pictures is that the critical roles of evanescent lateral modes and their contributory attributes are totally missed.

Our objective here is to clearly differentiate the effect of guided-mode resonance reflection and Mie resonance in a simple lattice built with isolated particles. We treat a 1D cylindrical-rod lattice providing rigorously-computed maps of zero-order reflectance $R_0$ in wavelength ($\lambda$) and period ($\Lambda$) for silicon nitride (n=2) and silicon (n=3.5) holding the rod diameter constant at D=250 nm throughout. We begin by showing that, in general, there is no connection between isolated Mie resonance and guided-mode lattice resonance. Then, by homogenizing the lattice with effective medium theory, we show a strong correlation between the $R_0=1$ resonance loci and spectral loci of the lateral modes belonging to the equivalent film. We compare the local field profiles in isolated rods with those of the lattice by numerical computations. We find that when the lateral leaky-mode field profiles approach the structure of the isolated-particle Mie field profiles, the resonance locus bends towards the Mie resonance wavelength. The interesting fact that the lattice fields "remember" the isolated particle fields is referred here as "Mie modal memory." We study the preservation and erasure properties of this memory effect. This work differs from a prior contribution in that in Ko et al.[20] perfect reflectance was retained as the Mie cavity was destroyed whereas here the cavity is retained. Thus, the present work provides a new alternate view that is straightforward in its interpretation.

## Mie resonance and lattice resonance: Distinction

Figure 1 illustrates models and spectra pertaining to Mie resonance in isolated particles and guided-mode resonance in periodic lattices. The model particle chosen is an infinite circular cylinder with diameter D and refractive index n placed in air (Fig. 1a). The lattice is an array of similar particles with period $\Lambda$ (Fig. 1d). The lattice operates in the subwavelength regime such that only the zero-order reflectance ($R_0$) and zero-order transmittance ($T_0$) are shown in Fig. 1d. The illuminating plane wave is at normal incidence with wavenumber $k_i$. As usual in diffraction and waveguide optics, we define TE and TM polarization state as electric field parallel and perpendicular to the particle axis. Figure 1b presents the total scattering cross section (TSCS) of a single particle with D = 250 nm and refractive index n = 2. Similarly, Fig. 1c provides the TCSC for n=3.5. On account of the cylinder geometry, we label the Mie resonance field configuration in terms of azimuthal mode number ($j$) and radial mode number ($l$) as $TE_M(j,l)$ or $TM_M(j,l)$.[22] In Fig. 1c, $TE_M(1,1)$ and $TM_M(0,1)$ are located at $\lambda = 1.179$ μm and 1.173 μm, respectively, or close to each other. Obviously, the TE-polarized TSCS exceeds the TM-polarized TSCS because TM light encounters Brewster conditions at the cylinder surface resulting in lower TM reflection and less effective scattering. Figure 1e presents $R_0$ spectra under TE and TM polarization for n=2. The resonance peaks are labeled $TE_L(m,v)$ or $TM_L(m,v)$ where $m$ denotes the evanescent diffraction order that generates the resonance and $v$ marks the corresponding classic waveguide mode. Figure 1f shows zero-order reflectance for the case of n=3.5. Under guided-mode lattice resonance in Figs. 1e and 1f, we see that $R_0=1$ for both polarization states at the respective resonance wavelengths. In his example, with a period chosen arbitrarily, there is no correlation between the Mie resonance wavelengths and the GMR wavelengths. This is because there is no causal relationship between the condition $R_0=1$ and Mie resonance which is one of the main points of this paper.

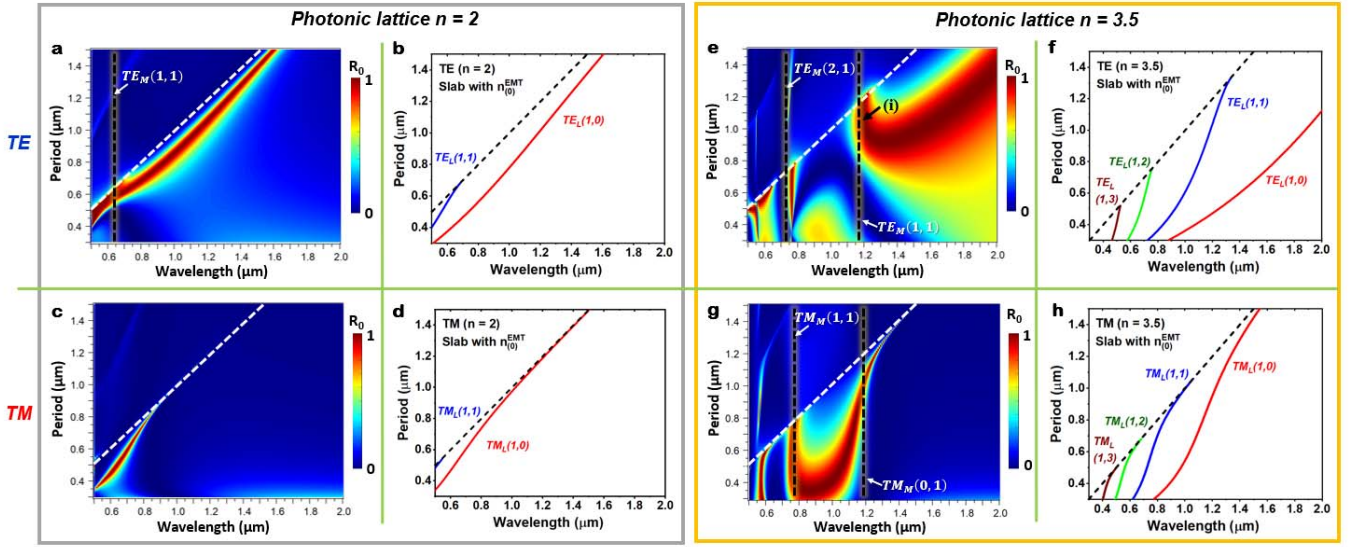

**Fig. 2 | Perfect reflection bands generated by resonant photonic lattices and their association with effective-medium mode loci.** Displayed are wavelength–period ($\lambda$-$\Lambda$) zero-order reflectance ($R_0$) color maps for cylinder arrays with D=250 nm as well as $\lambda$-$\Lambda$ modal curves for equivalent slab waveguides where the full Rytov EMT refractive index ($n_{(0)}^{EMT}$) is used. **a-d**, Resonance maps and modal curves for both polarization states for n=2. **e-h**, Resonance maps and modal curves for both polarization states for n=3.5. In the figures, the subwavelength regime resides below the dashed lines marking the Rayleigh wavelength $\lambda=\Lambda$. In the $R_0$ maps, 100% reflectance is featured by dark red color.

## Perfect reflection: Correlation with lateral modes

In Fig. 2, the zero-order $\lambda$-$\Lambda$ reflectance maps are computed with rigorous coupled-wave analysis (RCWA)[23]. The mode loci pertinent to the homogenized lattice are computed with effective medium theory (EMT) and waveguide theory. We apply the full Rytov formalism [24, 25] to extract the zero-order EMT ($n_{(0)}^{EMT}$) and use it to model the slab waveguide. We note that this effective index is a function of $\lambda$, $\Lambda$, and period-dependent fill factor F($\Lambda$). Finally, we solve the eigenproblem such that the propagation constant is the wavevector of first-order diffraction ($\beta = 2\pi/\Lambda$). The details of this calculation are explained in Supplementary Information.

Figure 2a provides a zero-order TE reflectance map for the lattice under study with n=2 corresponding approximately to $Si_3N_4$. In the wavelength range shown, there appears one Mie resonance at $\lambda=0.63$ μm (as seen in Fig. 1b) within the perfect-reflectance locus in the figure. There ensues no variation of any reflectance features at this point, demonstrating that $R_0=1$ holds at Mie resonance as well as away from it. The high-reflectance region borders the Rayleigh line $\lambda=\Lambda$ for part of the way and then diverges from it. The reason for this is that for the smaller periods, the particles (D=250 nm fixed) are relatively close together, forming a lattice with a substantial effective value of refractive index. This particle density thus allows the effective lattice to support two lateral leaky modes, namely the $TE_0$ and $TE_1$ modes. The $TE_1$ mode appears at shorter wavelengths and thus resides near the Rayleigh line as seen in Fig. 2a. As the period increases and reaches a value of $\Lambda \sim 750$ nm, the effective index drops sufficiently to cut the $TE_1$ mode off, accounting for the variation in the locus. Thereafter, as $\Lambda$ increases, the resonance proceeds on the $TE_0$ mode alone on an increasingly sparse lattice; we recall that the fundamental waveguide mode is never cut off. These arguments are well supported by Fig. 2b showing the mode loci of the homogenized lattice. There is quantitative agreement between the locus of the homogenized $TE_1$ mode labeled $TE_L(1,1)$ and the $R_0=1$ locus. For the fundamental mode $TE_L(1,0)$, there is good quantitative agreement for $\Lambda>0.7$ μm.

The case for TM polarization with n=2 is depicted in Fig. 2c. According to Fig. 1b, there is no Mie resonance within the perfect reflection region. Because TM effective index is lower than the TE index for the same lattice, only the fundamental $TM_L(1,0)$ mode survives here. There is excellent agreement with the mode locus in Fig. 2d. Because of the low effective index, the mode locus approaches the Rayleigh line more rapidly than in the TE case as the period increases.

If we plot analogous resonance maps for increasing rod refractive index, we see that the slanted-V type locus arising for n=2 in Fig. 2a gradually morphs into the locus for n=3.5 approximating Si in Fig. 2e with similar modal content. For this high value of refractive index, the homogenization is less accurate and the modal lines in Fig. 2f only qualitatively resemble the full numerical maps in Fig. 2e. Nevertheless, Fig. 2f marks the approximate cutoff of $TE_L(1,1)$ and shows the mild bow shape of the $TE_L(1,0)$ locus. It also shows the near vertical nature of the $TE_L(1,2)$ locus seen in Fig. 2e near wavelength of 0.8 μm. Similar comments apply to the TM case in Figs. 2g and 2h. With increasing index n, the $R_0=1$ locus morphs into the final shape in Fig. 2g. The mode line for $TM_L(1,0)$ overlaps a part of the reflectance locus in Fig. 2g from ~0.8 to ~1.2 μm wavelength. The mode line for $TM_L(1,1)$ approximates the vertical locus near 0.8 μm.

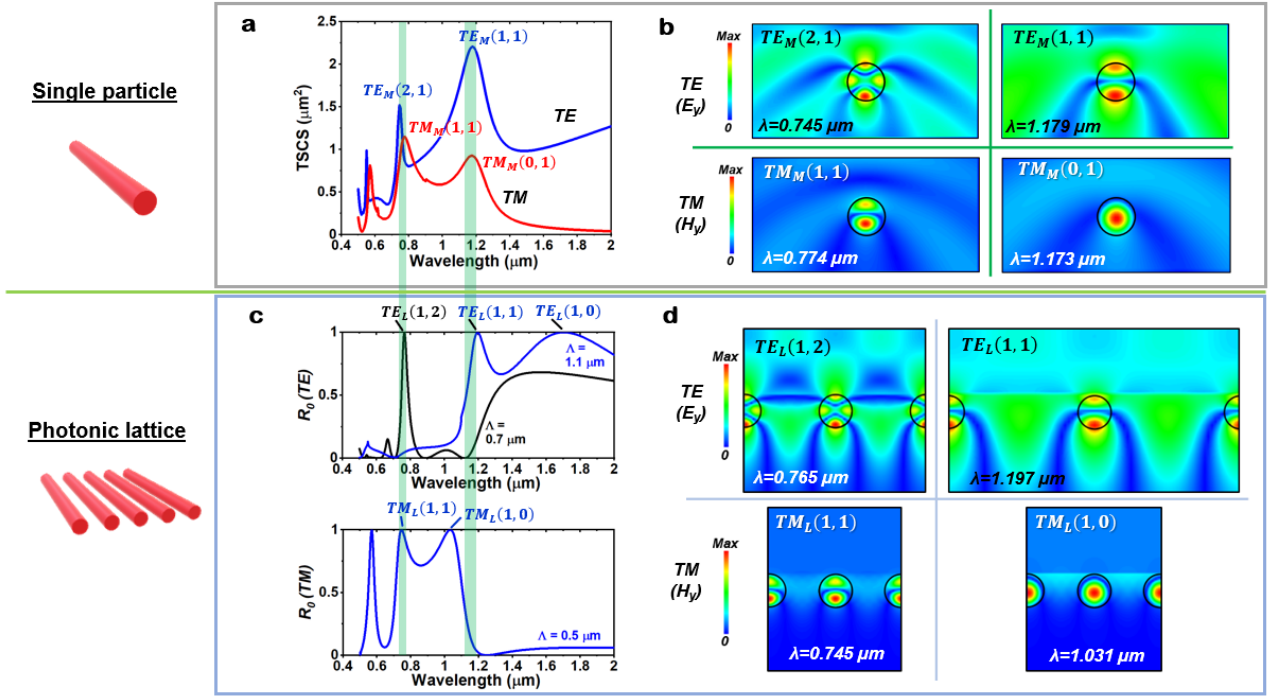

**Fig. 3 | Quantification of local/lateral mode matching.** We model a single circular cylinder with D=250 nm and n=3.5 and a corresponding lattice. **a**, Total scattering cross section (TSCS) spectra in TE and TM polarized light. Mie resonance peaks are labeled $TE_M(j,l)$ or $TM_M(j,l)$ by azimuthal mode number (*j*) and radial mode number (*l*). **b**, Electric and magnetic field profiles at Mie resonance wavelengths corresponding to **a**. E and H indicate the amplitudes of electric and magnetic fields. **c**, Photonic lattice spectra $R_0(\lambda)$ at values of Λ chosen to match overlapping Mie/lattice resonance locations in Fig. 2e for TE polarization and Fig. 2g for TM polarization. The guided-mode resonance peaks are labeled as $TE_L(m,v)$ or $TM_L(m,v)$ with *m* denoting the diffraction order and *v* the waveguide mode. **d**, E and H profiles at lattice resonance points corresponding to **c**. Comparing **b** and **c** verifies the local/lateral mode matching at these (λ,Λ) coordinates.

**Lateral/local mode matching**

Within the spectral range covered in Fig. 2e, there appear single-particle Mie resonances marked by vertical lines labeled $TE_M(1,1)$ and $TE_M(2,1)$ to identify the type of Mie modal profile. Similarly, in Fig. 2g, Mie resonance wavelengths are noted and labeled $TM_M(0,1)$ and $TM_M(1,1)$ for TM polarized single-particle Mie modes. It is notable that the perfect-reflectance $R_0=1$ loci bend towards these spectral locations such that there is strong correlation with the individual-particle resonance wavelengths and the lattice-resonance wavelengths at these locations. We explain this physical manifestation by spatial field matching between the Mie modes and the lateral modes generating the resonance. As the individual cylinders possess characteristic Mie resonance field profiles at the Mie resonance wavelengths, the lateral Bloch modes must match those at least approximately at these specific spectral (λ,Λ) coordinates.

To confirm this idea, we compare TSCS (λ) and $R_0(\lambda)$ spectra and attendant field profiles at values of Λ chosen to match overlapping Mie/lattice resonance locations in Figs. 2e and 2g. In Fig. 3a, Mie modes $TE_M(1,1)$ and $TM_M(0,1)$ occur at λ = 1.179 μm and 1.173 μm. At higher energy states, $TE_M(2,1)$ and $TM_M(1,1)$ are found at λ = 0.745 μm and 0.774 μm. With periods chosen for resonance coincidence to the extent possible, $R_0(\lambda)$ lattice spectra are displayed in Fig. 3c and compared to the Mie resonance wavelengths in Fig. 3a with vertical lines. For Λ=1.1 μm, the $TE_L(1,1)$ locates near $TE_M(1,1)$ and $TE_L(1,2)$ is closely matched to $TE_M(2,1)$ at Λ=0.7 μm. Similarly, $TM_L(1,1)$ is close to $TM_M(1,1)$ at Λ=0.5 μm. We now compare the localized field structures in the cylindrical single particles in Fig. 3b with the fields residing in the photonic lattice in Fig. 3d. We see that the resonant-lattice field patterns approximate the single-particle fields with good qualitative agreement. This is in spite of the fact that the guided-mode resonance wavelengths differ somewhat from the exact Mie resonance wavelengths as quantified in Figs. 3b and 3d. This wavelength difference is reasonable because of the geometric difference of the two physical arrangements. In the lattice, at resonance, there are contradirectional leaky Bloch modes interacting with the particles in addition to the incident wave in stark contrast with the single-particle case. The evanescent-wave-excited lateral modes interacting with the incident wave generate the perfect reflection with the approximate mode matching shown here bending the loci towards the Mie resonance wavelengths as illustrated in Figs. 2e and 2g. We conjecture that the mode-matching principle set forth here is general and will apply to any dielectric resonant optical lattice independent of the shape of the building block particles constituting the array.

## Mie modal memory: Preservation and erasure

Figure 3 demonstrates preservation of Mie resonance signature when the lattice resonance wavelength approximates the Mie resonance wavelength. Thus, whereas the lattice supports counterpropagating Bloch modes forming a standing wave, there appear field patterns reminiscent of the single-particle Mie resonance fields. Thereby, the photonic lattice acts as a Mie modal memory. Here, we briefly investigate the robustness of this memory effect relative to perturbation of the Mie cavity with a continuum layer as illustrated in Fig. 4a. When the homogeneous layer thickness is $d_h$ = D/10, the particle cavity persists in large measure and the Mie signatures remain in the (λ,Λ) reflectance map to the degree quantified in Fig. 4a. As $d_h$ gradually increases, the Mie memory fades. For example, for $d_h$ = 0.3D, as can be seen in Fig. 4b, the resonant Mie memory is erased due to the destruction of the cavity. Perfect reflectance endures without any connection to Mie resonance; this map now follows the effective-medium mode loci in Fig. 2f more closely. To visualize the subtleties of the resonant memory effect, we characterize the localized fields at points marked (i)-(iii) in Fig. 2e and Figs. 4a and 4b respectively. Figures 4c-e illustrate the transition of the resonant memory. First, in the discrete lattice of Fig. 2e at point (i), the perfect reflection band resides near the Mie resonance line. The lattice mode $TE_L$ (1,1) and the Mie mode $TE_M$ (1,1) have similar signatures as summarized in Fig. 4c. Figure 4d shows the conditions at point (ii). There, the Mie signature is largely retained with an additional field concentration appearing in the thin sublayer. However, at point (iii) in Fig. 4e, the thicker sublayer destroys the cavity thus erasing the Mie memory. The localized mode field merges into the sublayer showing a characteristic standing-wave profile.

The example in Fig. 4 pertains to TE polarization; similar effects are also observed in TM polarization.

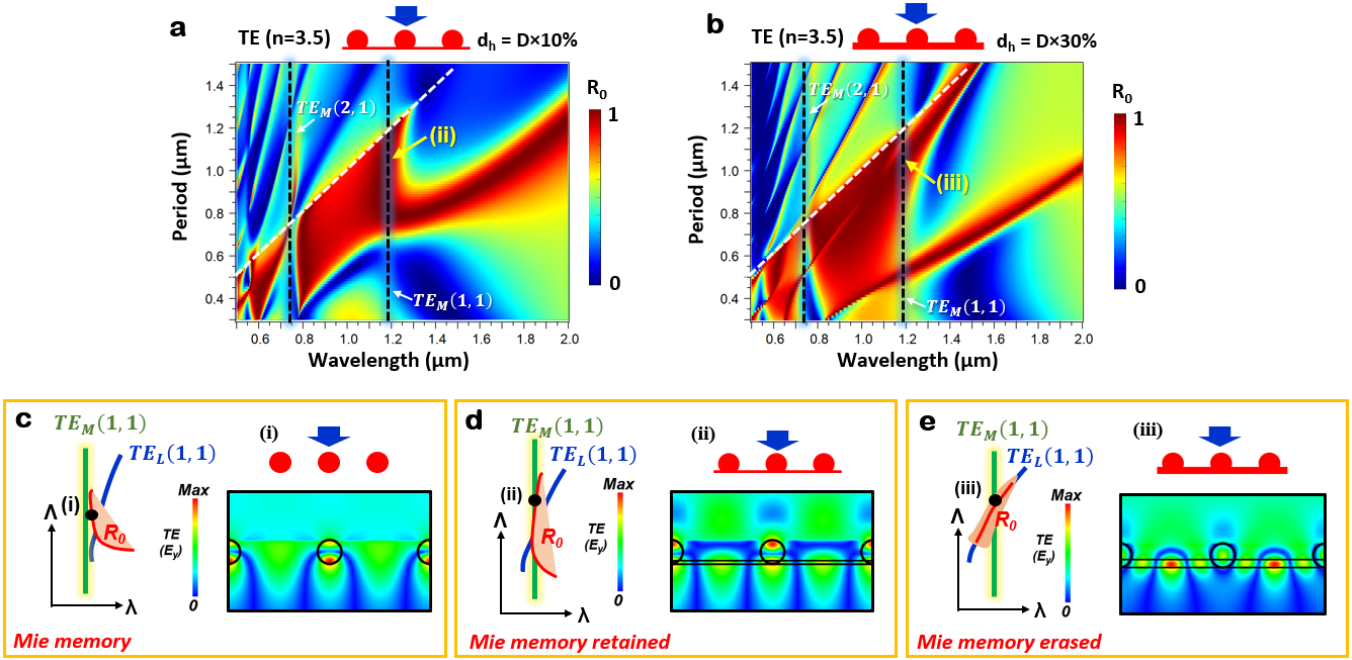

**Fig. 4 | Properties of Mie modal memory in TE polarization. a,** $R_0$ (λ,Λ) reflectance map with $d_h$ = D/10. **b,** $R_0$ (λ,Λ) reflectance map with $d_h$ = 0.3D. **c,** Original local fields and mode alignment at point (i). **b,** Fields and mode relationship at point (ii) upon perturbation with a thin sublayer. **c,** Local fields and mode alignment at point (iii) with Mie memory erased with a thick sublayer.

## Conclusions

In summary, we address the physics and origin of perfect reflection by resonant photonic lattices. The cause of this extraordinary effect is guided-mode resonance mediated by lateral Bloch modes excited by evanescent diffraction orders in the subwavelength regime. As we show clearly, Mie resonance is not causative in the perfect reflection by the lattice. Under conditions defined here, isolated-particle Mie resonance can, however, affect the spectral resonance location while not affecting the reflection efficiency.

For simplicity and clarity, we treat 1D arrays of dielectric cylinders with two representative materials with refractive indices n=2 and n=3.5. For arbitrary periods, single-particle resonance spectra and lattice-resonance spectra are uncorrelated. Maps of zero-order reflectance $R_0$ (λ,Λ) chart clear loci with $R_0$ =1 that are shown to associate strongly with simple waveguide modes supported by a homogenized effective-medium model of the lattice. For n=2, the agreement between the $R_0$ =1 loci and homogenized-slab mode loci is very good whereas for n=3.5 qualitative agreement is found. Moreover, for n=2, in TM polarization, there is perfect reflection for a range of (λ,Λ) coordinates even though there exists no Mie resonance in the region. For higher values of refractive index, the $R_0$ =1 loci bend towards the isolated-particle Mie resonance wavelengths. We explain this observation by spatial field matching between the Mie modes and the lateral modes inducing the guided-mode resonance. The individual cylinders possess characteristic local electric and magnetic field profiles at the Mie resonance wavelengths. We show that the lateral Bloch modes will match those profile shapes, at least approximately, at these specific spectral $(\lambda,\Lambda)$ coordinates. The interesting fact that the resonance lattice fields "remember" the isolated particle local fields is referred here as "Mie modal memory." Expectedly, this memory effect is strongest in lattices built with high-index materials. By connecting the individual lattice particles by an index-matched sublayer of sufficient thickness, the Mie memory can be erased. This is due to the destruction of the local Mie cavity. We find that perfect reflection survives the memory erasure with the resonance locus further approaching the effective-medium waveguide locus. The ideas presented here can be extended to two-dimensional lattices including sphere or pillar elements. The results presented here have potential to advance the field of nanophotonics, including metamaterials and metasurfaces, by solidifying the understanding of the physical basis of resonant photonic lattices.

## Methods

The isolated-cylinder scattering problem is solved by two-dimensional (2D) finite-difference time-domain (FDTD) methods utilizing commercial computational tools (Rsoft, FullWAVE module). In the simulation, we use an enclosed input source that launces a plane wave within the boundary surrounding the element.[26] To calculate reflectance spectra of periodic structures, we perform rigorous coupled-wave analysis (RCWA).[23] To establish connections between the high reflection bands and equivalent slab modes, we homogenize the lattice with effective-medium theory (EMT) models. The solution details are contained in Supplementary Information.

## Acknowledgements
This research is supported, in part, by the UT System Texas Nanoelectronics Research Superiority Award funded by the State of Texas Emerging Technology Fund as well as by the Texas Instruments Distinguished University Chair in Nanoelectronics endowment. Additional support is provided by the National Science Foundation (NSF) under award no. ECCS-1809143.

## Author contributions
Y.H.K. and N.R. performed RCWA and FDTD computations. H.H provided the Rytov EMT formalism. Y.H.K. solved the equivalent slab waveguide problem with full EMT. R.M. led the research and obtained funding. All authors contributed to the writing of the manuscript.

## Competing interests
The authors declare no competing interests.

# Perfectly-reflecting guided-mode-resonant photonic lattices possessing Mie modal memory

Yeong Hwan Ko[1], Nasrin Razmjooei[1], Hafez Hemmati and Robert Magnusson

Department of Electrical Engineering, University of Texas at Arlington, Arlington, Texas 76019, USA
Authors to whom correspondence should be addressed: magnusson@uta.edu
[1]Equal first authors

**Slab waveguide with Rytov EMT**

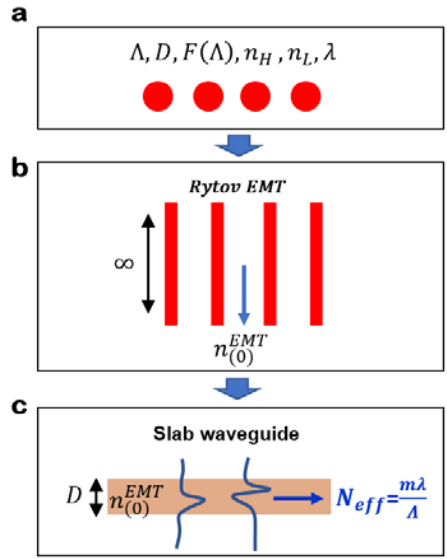

**Fig. S1 | Approximation of a slab waveguide with Rytov EMT. a**, Discrete array of cylinders where Λ, D, F, $n_H$ and $n_L$ are period, diameter, refractive indices of cylinder (n=2, 3.5) and air (n=1). **b**, Extraction of full effective refractive index ($n_{(0)}^{EMT}$). The grating parameter set {Λ, D, F, $n_H$, $n_L$} is applied in the Rytov formula. **c**, Modeling of a slab waveguide with $n_{(0)}^{EMT}$. For the Bloch mode with first-order diffraction (*m*=1), the eigenvalue problem is solved satisfying $\beta = 2\pi/\Lambda$.

To analyze the high reflection bands, we calculate GMR modal curves with an equivalent slab waveguide as explained in Fig. S1. In the beginning, we treat the discrete cylinder array with grating parameter set {Λ, D, F(Λ), $n_H$, $n_L$} where Λ, D, F, $n_H$ and $n_L$ are period, diameter, refractive indices of cylinder and background (Fig. S1a). Here, the effective fill factor F is given by $D\pi/4\Lambda$. Then, as illustrated in Fig. S1b, the grating parameter set is applied in the Rytov formulas for TE and TM polarization given by[1]

$$\sqrt{n_L^2 - (n_{TE}^{EMT})^2} \tan\left[\frac{\pi\Lambda}{\lambda}(1-F)\sqrt{n_L^2 - (n_{TE}^{EMT})^2}\right] = \sqrt{n_L^2 - (n_{TE}^{EMT})^2} \tan\left[\frac{\pi\Lambda}{\lambda}(1-F)\sqrt{n_L^2 - (n_{TE}^{EMT})^2}\right] \quad (1)$$

$$\frac{\sqrt{n_L^2 - (n_{TM}^{EMT})^2}}{n_L^2} \tan\left[\frac{\pi\Lambda}{\lambda}F\sqrt{n_H^2 - (n_{TM}^{EMT})^2}\right] = -\frac{\sqrt{n_L^2 - (n_{TM}^{EMT})^2}}{n_H^2} \tan\left[\frac{\pi\Lambda}{\lambda}F\sqrt{n_H^2 - (n_{TM}^{EMT})^2}\right] \quad (2)$$

Where we used the lowest-order solution $n_{(0)}^{EMT}(\Lambda, \lambda)$ by solving Eqs. (1) and (2). Using the extracted Rytov EMT, we solve the eigenvalue problem of the single slab waveguide as shown in Fig. S1c. For modeling, the single layer is homogenized by $n_{(0)}^{EMT}$. Then, the modal curves are calculated by solving eigenvalue equation for the effective slab waveguide as[2]

$$(\gamma_m D) \cdot tan[(\gamma_m D) + v\pi] = \sqrt{V^2 - (\gamma_m D)^2} \quad (v \text{ is even}) \quad (3)$$

$$(\gamma_m D) \cdot cot[(\gamma_m D) + v\pi] = -\sqrt{V^2 - (\gamma_m D)^2} \quad (v \text{ is odd}) \quad (4)$$

where the $V$-parameter and $\gamma_m$ are obtained by $V = 2\pi D\sqrt{(n_{TE}^{EMT})^2 - 1}/\lambda$ and $\gamma_m = 2\pi\sqrt{(n_{TE}^{EMT})^2 - (\lambda m/\Lambda)^2}/\lambda$. In the eigenvalue problem, we label the solution as $TE_L(m, v)$ when the GMR forms at the $m$th diffraction order coupled to the $v$th guided mode. In a similar way, we obtain the $TM_L(m, v)$ mode set by solving the TM slab waveguide problem.